# Unidirectional Localization of Photons

Hamidreza Ramezani[1], Pankaj Jha[2], Yuan Wang[2], Xiang Zhang[2]

[1] Department of Physics, The University of Texas Rio Grande Valley, Brownsville, TX 78520, USA
[2] NSF Nano-scale Science and Engineering Center (NSEC), University of California, Berkeley, 3112 Etcheverry Hall, Berkeley, CA 94720, USA

**Artificial defects embedded in periodic structures are important foundation for creating localized states with vast range of applications in condensed matter physics[1], photonics[2,3] and acoustics[4]. In photonics, localized states are extensively used to confine and manipulate photons[5–11]. Up to now, all the proposed localized states are reciprocal and restricted by time reversal symmetry. Consequently, localization is bidirectional and photons at the allowed passband in the otherwise forbidden stop band are confined irrespective of the direction of incident beam. In this report, by embedding a single defect in a one-dimensional spatiotemporally modulated photonic lattice, we demonstrate that it is possible to have localization of photon only in one direction. In a spatiotemporally modulated photonics lattice, a time dependent potential generates an effective magnetic biasing, which breaks the reciprocity[12]. Moreover, in such moving lattices the dispersion relation obtains a shift depending on the direction of effective magnetic biasing. A static defect synthesized in a temporally modulated lattice will generate a spatial localization of light in the bandgap. However, due to the shift of the bandgap the localization occurs in different frequencies depending on the direction of incident field. We envisage that this phenomenon might has impact not only in photonics but also other areas of physics and engineering such as condensed matter and acoustics, opens the doors for designing new types of devices such as non-reciprocal traps, sensors, unidirectional tunable filters, and might result in unconventional transports such as unidirectional lasing. Despite its applications, our proposal, namely a defect sate in a driven system, can be considered as a pedagogical example of Floquet problem with analytical solution.**

Formation of the band structure is an intriguing feature of the periodic systems. Many physical phenomena in the periodic systems are associated with different properties of their band structures. The concept of periodicity gives birth to the photonic crystals[13] where the dielectric constant in the structure is altered periodically. In a photonic crystal, guided modes are separated by the photonic bandgaps in dispersion relations. Photon propagation is forbidden or strongly suppressed in the photonic bandgap. Consequently, by engineering the electric permittivities one can achieve structures with all unusual band diagrams to guide and mould the flow of light[14,15]. Insertion of a defect in the periodic systems generates a high Q resonant mode in the gap which leads to spatial localization of wave functions in the vicinity of the defect[16,17]. The high Q modes are good candidates for designing photonic crystal lasers[5–7,18]. Embedding a pumped gain medium as a non-Hermitian defect in the photonic crystal results in the concentration and amplification of the light until it reaches to the threshold and starts to lase. Recent advancement in the non-Hermitian systems extended the concept of the photonic crystals to a new direction. Intriguing features are proposed such as reconfigurable Talbot effect[19], conical diffraction based on the exceptional point[20], unidirectional lasing[21,22] and unidirectional antilaser[23]. Nevertheless, all the aforementioned phenomena are restricted by reciprocity. Forwards and backward traveling fields observe the same structure and thus behave exactly the same. Therefore, band structure is symmetric and any unidirectional light propagation is prohibited.

Lattices with time symmetry ($\mathcal{T}$) or more precisely any symmetry that change wavevector $\vec{k} \to -\vec{k}$ does not support asymmetric band structure[24]. Consequently, transmitted field in such lattices is symmetric and independent from the input channel. Nevertheless, in recent years there is a demand to obtain asymmetric transport especially in miniaturized and compact systems[25–28]. Thus, one needs to break the reciprocity to obtain asymmetric transport.

Magnetic biasing is the most common technique to break the reciprocity[29]. In Faraday isolators, we use magnetic field to rotate the polarization of the light and eliminate the undesirable signal by means of polarizers[30]. In a similar fashion, a periodic stack of anisotropic dielectrics and gyrotoropic magnetic layers results in asymmetric band structures[24,31]. By incorporating gain mechanism in such periodic magneto-optical layered medium, unconventional and robust lasing modes with directional emission have been proposed[21].

More recently, one-way mode conversion in waveguides has been proposed by means of spatiotemporally modulated index of refraction[12]. Such asymmetric mode conversion results in magnetic free non-reciprocal transport[26]. Later this idea brought into acoustics[32,33], radio-frequency[34] and cold atoms[28]. It has been shown that temporal potentials are imitating a magnetic field which condemn the undesired signals[35–37].

Here we propose a non-reciprocal defect mode where at a specific frequency the field gets an exponential form around the defect only in one direction. Specifically, in a non-reciprocal localized mode, the unidirectional trapping of photons results in the unidirectional exponential accumulation of photons travelling in a specific direction. For a finite system size, such localized mode results in non-zero transmission in the bandgap. In the opposite direction and at the same frequency, photons end up in the bandgap and thus their propagation is forbidden. For a reasonably strong modulation, we show that one can obtain an interesting situation, where in one direction photons gets trapped, namely localized, while in the opposite direction and at the same frequency the photons are in the passband with scattering state feature. Particularly in a scattering state, unlike the localized mode, the field does not have exponential form and has a different distribution than the localized mode. Finally, we show the frequency shift of the defect state is linearly proportional to the detuning similar to the Zeeman effect. The non-reciprocal defect state can filter the unwanted frequencies in the bandgap and transmit the defect mode signal. By changing the detuning one can tune the filtering frequency in a nonreciprocal manner.

In order to realize a unidirectional localized mode, as depicted schematically in Fig. (1), we embed a defect in a periodic spatiotemporally modulated 1D lattice. Although our proposal is general and can be implemented in different wave-base systems, we consider a periodic photonic lattice generated in a three level electromagnetically induced transparency (EIT) medium.

The EIT medium composed of $10^{19}$Rb atoms, which are in a cell of total length $\approx$ 2 mm that has two parts separated by a SiN dielectric membrane with length 88.6 nm and refractive index $n = 2.2 + 10^{-4}i$. The modulated refractive index of the EIT medium can be imposed by two detuned counter-propagating fields with frequencies $\omega_1, \omega_2$ and relative detuning $\delta = \omega_2 - \omega_1$. Moving the periodic intensity of the strong standing coupling field, namely $\delta \neq 0$, creates the time dependent modulation of the refractive index of a weak probe[38]. The membrane in the cell acts as a defect inside the photonics crystal formed by the EIT medium. We can drive (see supplementary materials) the Schrödinger like coupled mode equation for the weak forward and backward traveling fields generated by probe in the spatiotemporal modulated medium

$$i\frac{d}{dz}\begin{pmatrix}E_f\\E_b\end{pmatrix} = H(z)\begin{pmatrix}E_f\\E_b\end{pmatrix}, H(z) = \begin{pmatrix}\kappa_{11}(\omega_f) & \kappa_{12}(\omega_b)e^{-i\Delta kz}\\\kappa_{21}(\omega_f)e^{i\Delta kz} & \kappa_{22}(\omega_b)\end{pmatrix}. \quad (1)$$

The off diagonal terms in $H(z)$ mix the waves while the diagonal ones are attenuation coefficients, associated with the probe field with frequency $\omega_f$ propagating in the $z$ direction. Notice that the field propagation in a time dependent spatially modulated waveguide system is given by a Hamiltonian that has a form similar to $H(z)$[12]. Inside the defect, Helmholtz equation describes field propagation

$$\frac{d^2E}{dz^2} + \left(\frac{n\omega}{c}\right)^2 E = 0 \qquad (2)$$

where $E$ is the electric field inside the membrane with frequency $\omega$. Direction of propagation is considered to be $z$ and $c$ is the speed of light.

The three level EIT medium in figure (1) prohibits transition from level $|b\rangle$ and $|c\rangle$ due to the symmetry while transitions from $|a\rangle$ to $|c\rangle$ and vice versa is possible by the counter-propagating fields. Transition between $|a\rangle$ and $|b\rangle$ occurs due to the existence of the probe field with frequency $\omega_f$. The moving photonic crystal can reflect the probe field at frequency[28]

$$\omega_b = \omega_f \pm \delta. \qquad (3)$$

In Equation (3) the plus sign is associated with the probe field incident from the left and minus sign is associated with the probe filed incident from the right. Frequency of the transmitted signal does not obtain any phase shift as there is no wave mixing process to generate other frequencies.

Transfer matrix method can be used to calculate the transmission and reflection in the above moving photonic crystal. Figure (2a) depicts the transmission and reflection coefficients vs. probe detuning $\Delta_f \equiv \omega_{ab} - \omega_f$ in the absence of the membrane and the detuning $\delta$ and with $\omega_{ab} \approx 2414191.334$ GHz. The spatial periodicity of the dielectric constant of the photonic crystal generates Bragg reflection where photon propagation is forbidden in a window known as bandgap. Thus, in the bandgap transmission coefficient drops to zero. Considering the intrinsic losses in the system at the photonic bandgap the reflection plus absorption become one. By inserting the membrane in the cell, Fig.(2b), a defect mode with nonzero transmission appears in the bandgap at $\Delta_f \approx -0.12$ GHz. Transmission peak of the defect mode is reciprocal and *degenerate,* namely, irrespective of the direction of the incident field the transmission peak occurs at the same frequency. When we introduce a nonzero detuning, i.e. the permittivity of the photonic lattice becomes both space and time dependent, degeneracy breaks and the defect state frequency starts splitting for the left and right incident beams. Neglecting the higher quasi energies, we plotted the left incident transmission and reflection in the figure (2c) for $\delta = 0.015$MHz. In this case the defect mode appears at the probe detuning $\Delta_f \approx -0.37$GHz. On the other hand, in figure (2d) we observe that for the right incident field the defect mode appears $\Delta_f \approx -0.28$GHz.

Transmission peak at the defect mode has a sharp feature in the absence of the losses. At the defect mode, photons are trapped and the electric field is localized around the membrane. Specifically, the electric field envelope decays exponentially as we move away from the defect. This contrasts with the resonant peaks at the scattering states where the field is distributed all over the photonic crystal. In our photonic lattice a comparison between Fig.(2b) and Fig. (2c,d) shows that due to the time dependent modulation the position and width of the bandgap window is varying by changing the detuning and at the same time it affects the position of the localized modes. To distinguish the localized mode from the scattering states one should plot the field distribution for the localized mode. As long as the field has an exponential form we have the trapping of photons. However, eventually for very strong detuning the mode will merge completely to the band and its associated field distribution will not have an exponential shape. Specifically, one can use the detuning to tune the mode from being completely localized to a non-localized one.

To distinguish between scattering state, localized state and a state in the bandgap, in figures (3a-i) we plotted the field distribution in the photonic crystal for different detuning and at different probe detuning. Specifically, in Figs.(3a-c) we plotted the field at $\varDelta_f = -0.1179$GHz (localized state with exponential form), $\varDelta_f = 0.2162$GHz (scattering state), and $\varDelta_f = -0.5$GHz (a state in the bandgap) for $\delta = 0$. We clearly observe the difference in the field distribution at each case. To compare the field distribution for non-zero detuning we plot the field for the left incident and the right incident beam at $\varDelta_f = -0.36653$GHz, $\varDelta_f = -0.276522$GHz, and $\varDelta_f = -0.155$ GHz for $\delta = 0.015$MHz in Figs.(3d-f) and Figs.(3g-i), respectively. In all cases for the defect mode the exponential decay around the membrane is clearly observed. It is interesting that for $\varDelta_f = -0.36653$GHz the photons coming from the left side will be localized while for the same photons coming from the right they will be in the bandgap and get reflected. However, for $\varDelta_f = -0.276522$GHz the photons coming from the right will be localized and photons with the same frequency coming from the left side will be in the band and form a scattering state with finite transmission and reflection.

As discussed previously the nonzero detuning between the counter propagating fields splits the left incident and right incident defect modes. The splitting is linear with respect to the detuning similar to the Zeeman effect where a magnetic field splits the degenerate modes. This similarity between time dependent potentials and magnetic field is the basic principle behind the breaking of the Lorentz reciprocity. Magnetic field is the most common approach to break the reciprocity. However, to our knowledge there is no report on the existence of unidirectional localized mode based on magnetic effects. We mentioned earlier that the frequency of the localized mode is linearly dependent on the detuning. In Figure (4) we numerically calculated the changes of the frequency of the localized modes for the left and right incident fields versus the detuning. A linear fitting shows that our anticipation is correct and it behaves linearly in similar fashion as Zeeman effect. This indicates that one can consider an effective magnetic field in the system which breaks the reciprocity.

Any type of isolator based on magnetic field or time dependent modulation needs an absorbing and/or filtering channel to remove the undesired signal. Otherwise, in the absence of the filtering channel, the undesired field will be able to pass through the isolator after several forward and backward propagations. Our proposal is not distinct in this sense. However, in our case the undesired signal is in the gap and needs to travel many times to be able to pass through our proposed the isolator. We mentioned earlier that naturally, there are some intrinsic distributed losses in our optical system. These distributed losses are the main reason that in Figure (2) we do not have the conservation relation $T + R = 1$. Consequently, the losses will eliminate the undesired signal in our photonic lattice. Thus, our proposal is more compact with respect to the other isolators.

In Conclusion, we have shown that by embedding a defect in spatiotemporally periodic modulated photonic lattice one can achieve a unidirectional defect mode where the photons propagating in one direction become localized and get trapped in the bandgap. Whilst in the opposite direction photons with the same frequency get reflected or transmitted depending on the position of the mode in the bandgap window. This contrasts with the periodic spatial modulated case where a defect generates a reciprocal defect mode. Moreover, we showed that the position of the defect mode is tunable and depends on the strength of the temporal modulation. Specifically, the position of the defect mode linearly changes with respect to the temporal modulation. Our analytical results not only have important application in optics and photonics, but also can be consider as a pedagogical example of solvable defect in a Floquet system.

Acknowledgment: H. R acknowledge funding support from the UT system under the Valley STAR award. H.R. conceived the idea, performed analytical and numerical calculation associated with the

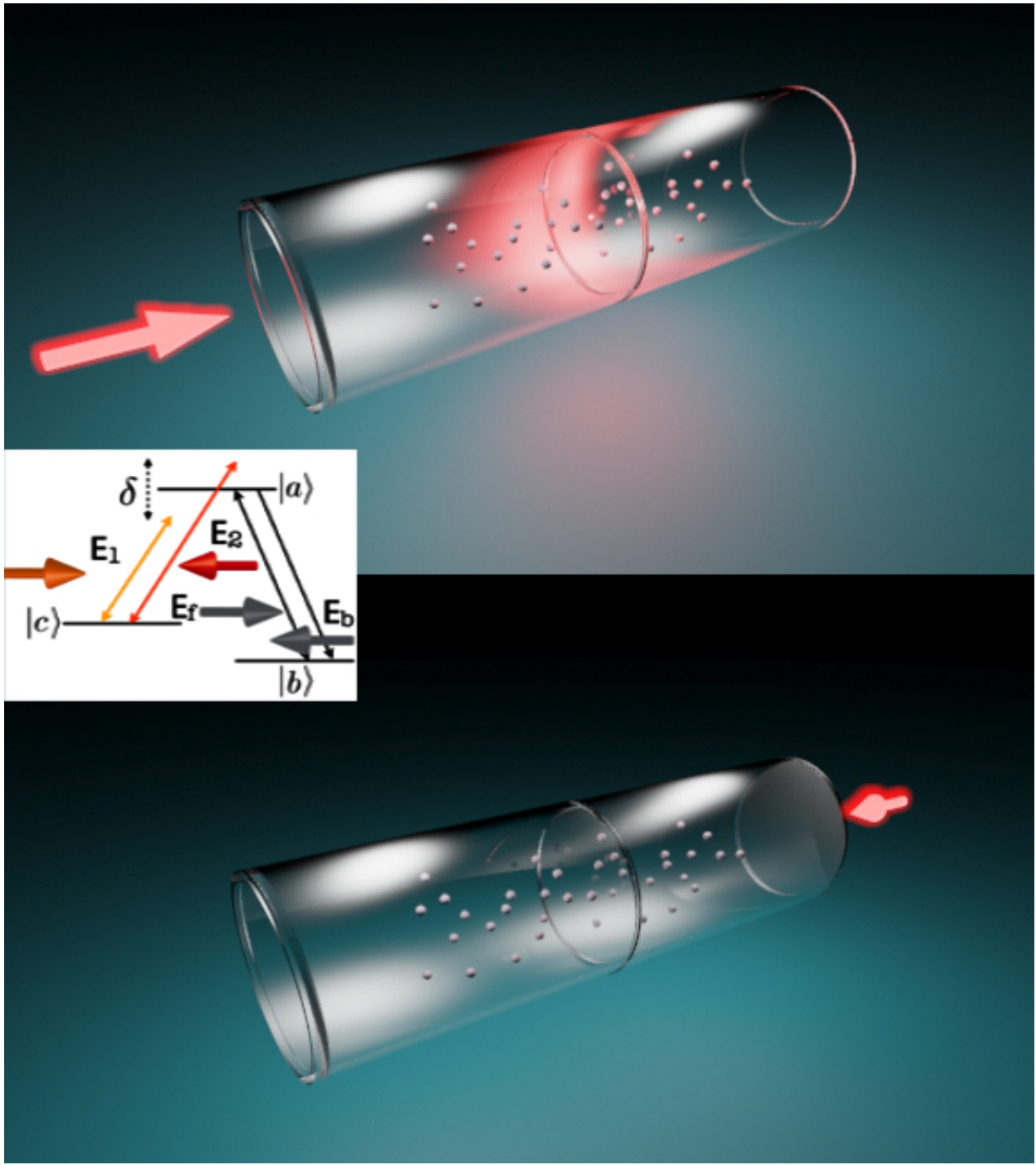

*Figure 1* Unidirectional localized mode in a spatiotemporal modulated photonic crystal with a static defect membrane at the center of the crystal. The photonic crystal is formed from a driven Rb atom cell (Λ type three level system) with a standing wave field with detuning $\delta$ between the two components. The upper figure shows schematically that the incident beam is localized at the defect membrane in the center of the cell while the right incident beam (lower figure) is not localized. The level structure of the EIT medium is represented in the inset of the lower panel.

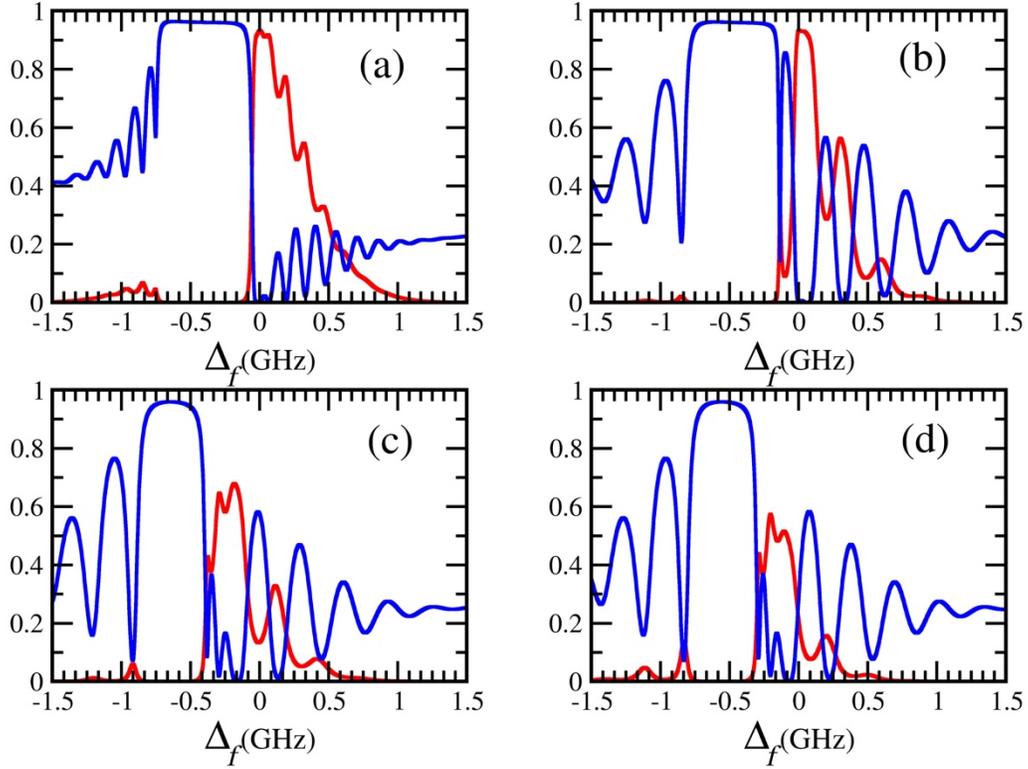

Figure 2 (a) Transmission (red) and reflection (blue) from the static photonic crystal ($\delta = 0$) with no defect. In the bandgap transmission reaches t zero. (b) Transmission (red) and reflection (blue) for the static photonic crystal ($\delta = 0$) with a defect in the middle of the lattice. The defect mode is appeared at the $\Delta_f \approx -0.13$(GHz). The left and right incident beam has the same transmission and reflection curves. (c) Left incident transmission (red) and reflection (blue) for the time modulated photonic crystal ($\delta = 0.015$ MHz) with a membrane defect in the middle of the lattice. The defect mode is appeared at the $\Delta_f \approx -0.38$(GHz). (c) Right incident transmission (red) and reflection (blue) for the time modulated photonic crystal ($\delta = 0.015$ MHz) with a membrane defect in the middle of the lattice. The defect mode is appeared at the $\Delta_f \approx -0.29$(GHz). Comparison between the (c) and (d) shows that the frequency for which we have the defect mode for left incident beam, the right incident beam observes the bandgap and has zero transmission. On the other hand, for the frequency for which we have a defect mode for the right incident beam the left incident beam observes the bandpass window.

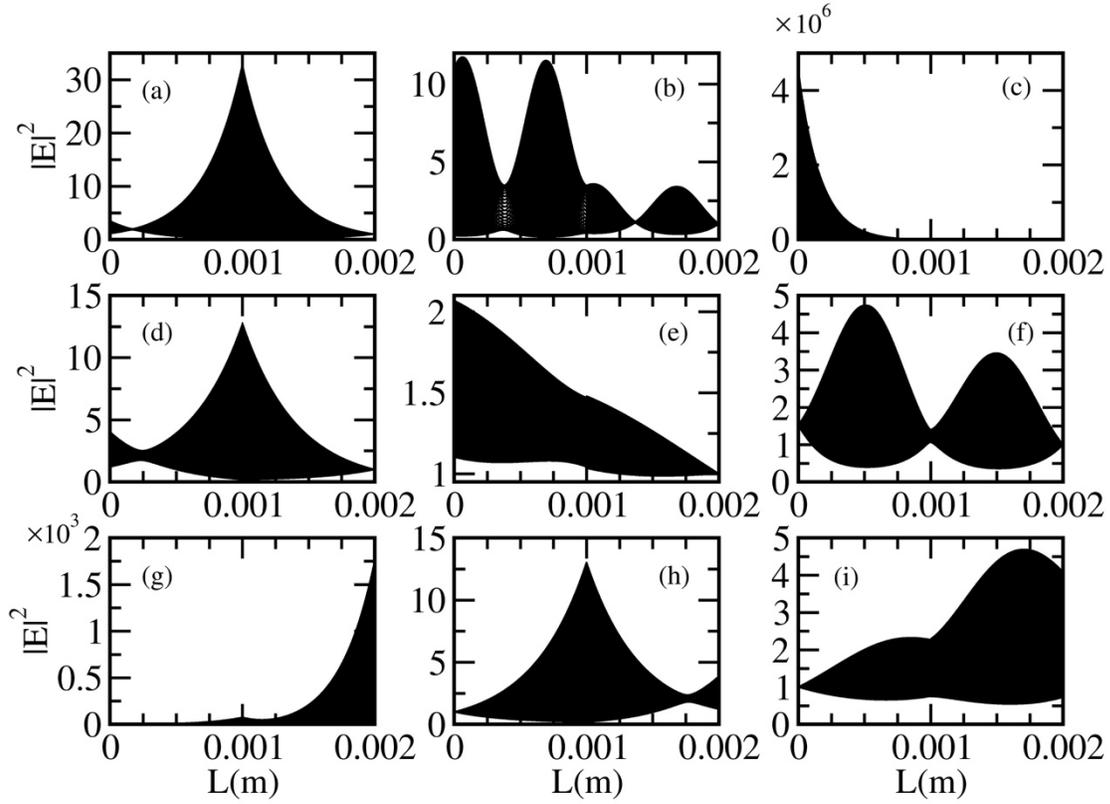

Figure 3 (a-c) Distribution of the field intensity for the zero detuning at the (a) defect mode, in the (b) passband window, and in the (c) gap. (d-f) Distribution of the field intensity for left incident and $\delta = 0.015$MHz at the defect mode $\Delta_f \approx -0.38$GHz (d), in the passband window $\Delta_f \approx -0.29$GHz (e), and at $\Delta_f \approx -0.155$GHz (f). (g-i) Distribution of the field intensity for the right incident and $\delta = 0.015$MHz at the gap for the right incident $\Delta_f \approx -0.38$GHz (d), at the defect mode $\Delta_f \approx -0.29$GHz (e), and at $\Delta_f \approx -0.155$GHz. Notice that the defect mode of the left incident beam is located at the gap for the right incident beam while the defect mode of the right incident beam is located at the passband and has scattering feature.

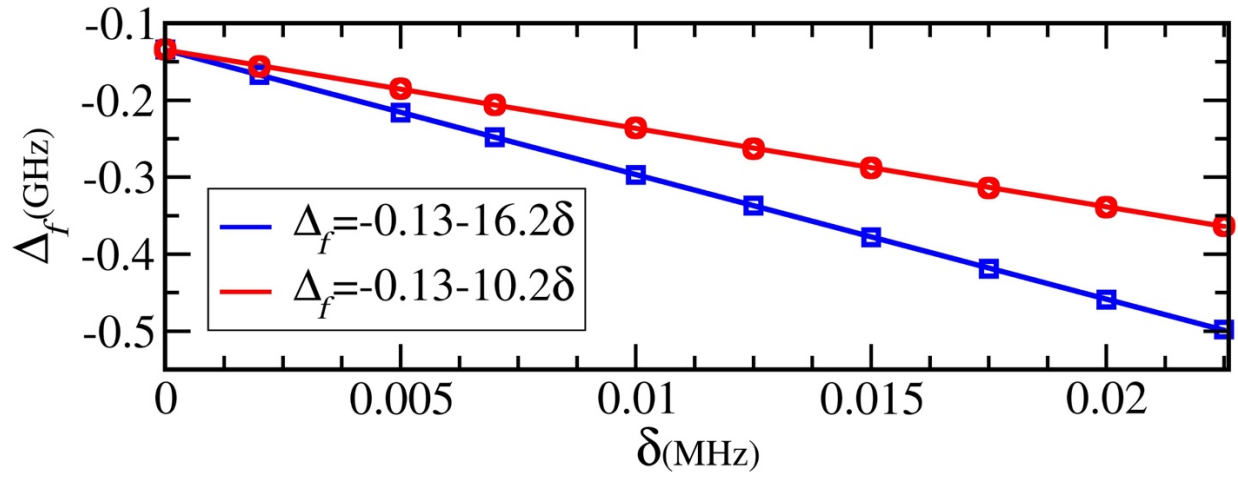

*Figure 4* Position of the defect mode vs. the detuning for the left incident beam (squares) and right incident beam (circles). A Linear fit is depicted by a continues line on top of the symbols. The splitting of the position of the modes shows a linear behavior similar to the Zeeman effect, showing the similarities between time dependent modulated lattice and a magnetic biasing.